\def\J#1#2#3#4{#1~{\bf #2},~#3~(#4).}
\def\K#1#2#3#4{#1~{\bf #2},~#3~(#4)}
\def\JAC{J. Appl. Crystallogr.~}
\def\JCG{J. Cryst. Growth}
\def\JJAP{Jpn. J. Appl. Phys.~}
\def\JAP{J. Appl. Phys.}

\def\PR{Phys. Rev.~}
\def\JPCM{J. Phys.: Condens. Matter~}
\def\PRL{Phys. Rev. Lett.~}
\def\PRX{Phys. Rev. X~}
\def\APL{Appl. Phys. Lett.~}
\def\JAP{J. Appl. Phys.~}

\def\PRB{Phys. Rev. B~}
\def\PRA{Phys. Rev. A~}

\def\DRM{Diamond Relat. Mater.~}

\def\S{Science~}
\def\N{Nature (London)~}
\def\Nmat{Nature Mater.~}
\def\Nphoton{Nature Photon.}
\def\Ncom{Nature Comm.}

\def\PNAS{Proc. Nat. Acad. Sci.}
\def\NL{Nano Lett.}

\def\APEX{Appl. Phys. Express}
\def\JAC{J. Appl. Crystallogr.}

\documentclass[twocolumn,showpacs,preprintnumbers,amsmath,amssymb]{revtex4} 
\usepackage{color}
\usepackage{graphicx}          

\usepackage{dcolumn}
\usepackage{bm}

\begin{document}

\title{Atomistic mechanism of perfect alignment of nitrogen-vacancy centers in diamond}

\author{Takehide Miyazaki$^{1,6}$, Yoshiyuki Miyamoto$^{1,6}$, Toshiharu Makino$^{2,6}$, Hiromitsu Kato$^{2,6}$, Satoshi Yamasaki$^{2,6}$, 
Takahiro Fukui$^{3}$, Yuki Doi$^{3}$, 
 Norio Tokuda$^{4}$, Mutsuko Hatano$^{5,6}$, and Norikazu Mizuochi$^{3,6}$}
\affiliation{
$^{1}$Nanosystem Research Institute, National Institute of Advanced Industrial Science and Technology, 
1-1-1 Umezono, Tsukuba, Ibaraki 305-8568, Japan, \\
$^{2}$Energy Technology Research Institute, National Institute of Advanced Industrial Science and Technology, 
1-1-1 Umezono, Tsukuba, Ibaraki 305-8568, Japan, \\
$^{3}$Graduate School of Engineering Science, Osaka University, Toyonaka, Osaka 560-8531, Japan, \\
$^{4}$Graduate School of Natural Science and Technology, Kanazawa University, Kanazawa 920-1192, Japan,\\
$^{5}$Department of Physical Electronics, Tokyo Institute of Technology, Meguro, Tokyo 152-8552, Japan, \\
$^{6}$CREST, Japan Science and Technology Agency, Kawaguchi, Saitama 332-0012, Japan.}
\email{takehide.miyazaki@aist.go.jp}

\begin{abstract}
Nitrogen-vacancy (NV) centers in diamond have attracted a great deal of attention because of their possible use in information processing and electromagnetic sensing technologies. We examined the atomistic generation mechanism for the NV defect aligned in the [111] direction of C(111) substrates. We found that  N is incorporated in the C bilayers during the lateral growth arising from a sequence of kink propagation along the step edge down to [$\bar{1}$$\bar{1}$2]. As a result, the atomic configuration with the N-atom lone-pair pointing in the [111] direction is formed, which causes preferential alignment of NVs. Our model is consistent with recent experimental data for perfect NV alignment in C(111) substrates.
\end{abstract}

\maketitle

Nitrogen-vacancy (NV) centers in diamond \cite{Doherty2013} have been recognized as representative examples of entangled $S$$=$1 systems in solid-state materials. 
Because of the externally controllable entanglement between the electronic and nuclear spin states  \cite{Jelezko2004,Dutt2007,Neumann2008} and excellent coherence properties in time  \cite{Balasubramanian2009} as well as space \cite{Bernien2013}, numerous technological applications of the NV centers have been demonstrated to date  \cite{Toyli2012,Neumann2010,Kucsko2013,Toyli2013,Maze2008,Balasubramanian2008,Grotz2012,Doi2014,Dolde2014,Hodges2013,Mizuochi2012,Lohrmann2011,Kurtsiefer2000,Fu2007,Sage2013,Igarashi2012,Steinert2013}.

\begin{figure}[b]
\begin{center}
\includegraphics[width=7.5cm,bb=0 0 606 297]{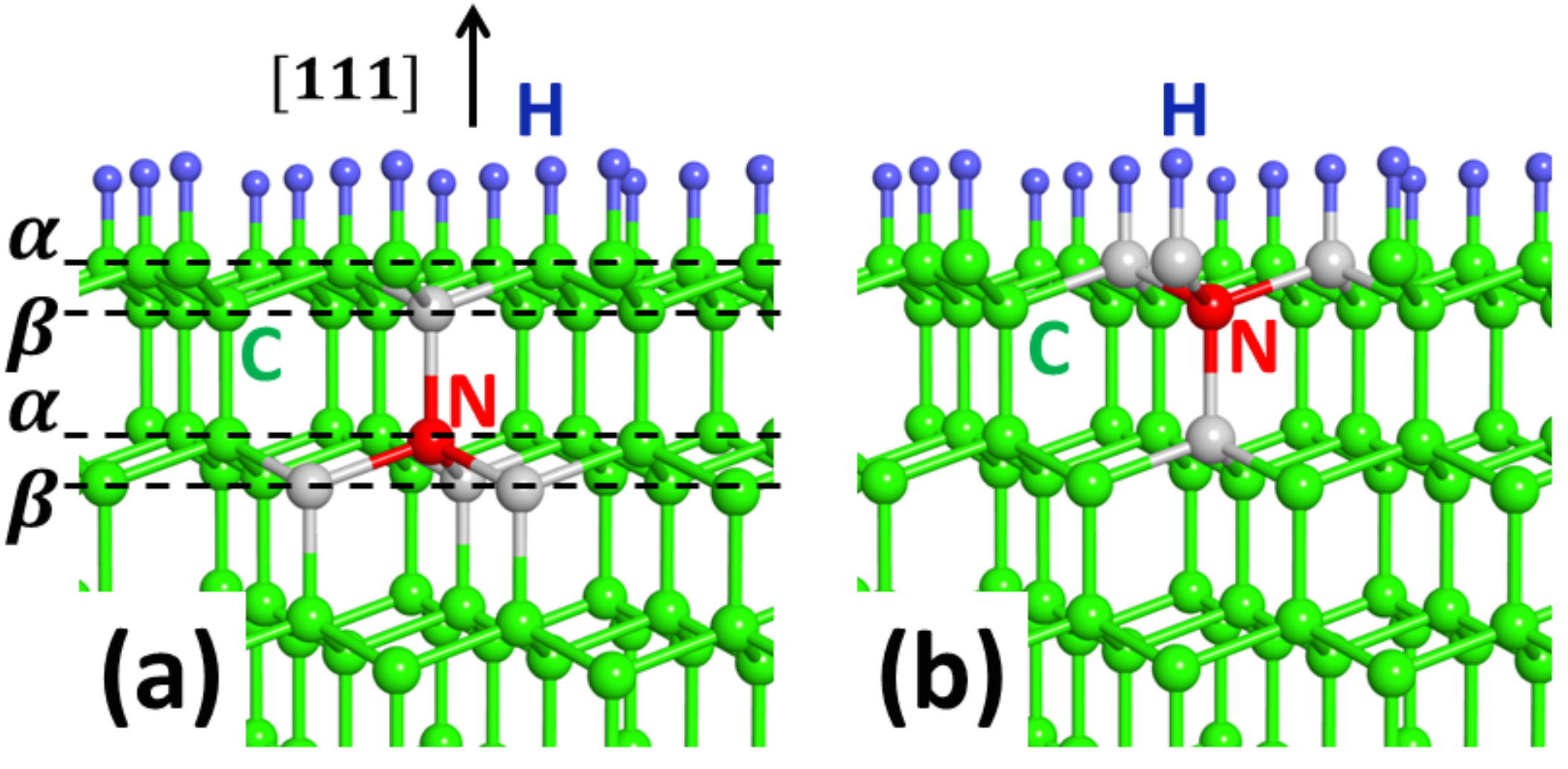}
\end{center}
\caption{NV center near the C(111):H surface. In panels (a) and (b), the gray balls adjacent to N illustrate  four possible  positions of V. 
(a) $\alpha$ and $\beta$ represent the two triangular sublattices in one bilayer. 
The $\alpha$ and $\beta$ layers for third bilayer are omitted for clarity. (a) N  in the $\alpha$ sublattice of the second bilayer. (b) N in the $\beta$ sublattice of the first bilayer.}
\label{f1}
\end{figure}

An NV center in a diamond crystal is composed of an N atom substituting for a carbon atom with an adjacent   carbon vacancy (V) sitting in various orientations relative to N.
For example, the NVs near the C(111) surface could have eight possible NV orientations, four with N in one sublattice in the bilayer [$\alpha$ layer, Fig. \ref{f1} (a)] and the remaining four with N in the other sublattice [$\beta$ layer, Fig. \ref{f1} (b)].

For practical applications, it is desirable to have the NV orientations aligned in one direction. In spin-dependent NV fluorescence experiments in C(001) samples grown by chemical vapor deposition (CVD), 
the magnetic field sensitivity was doubled relative to samples with random NV orientations \cite{Pham2012}. This was because most of NV centers were arranged in two of four possible orientations.
The partial alignment of the NV centers was also reported for CVD grown (110) substrates  \cite{Edmonds2012}. 

Recently, various groups \cite{Michl2014,Lesik2014,Fukui2014} have reported the nearly perfect alignment of the NV centers along the [111] direction in C(111) samples grown by CVD (94\%$\pm$2\% \cite{Michl2014}, $\sim$97\% \cite{Lesik2014}, and $\geq$99\% \cite{Fukui2014} relative to the total number of NVs generated). 
The formation mechanism was explained in terms of the step-flow growth \cite{Lesik2014} in a similar way to the (110) case \cite{Edmonds2012}. A first-principles study published prior to these experiments predicted that V and N are in the first and second layers, respectively [Fig. \ref{f1} (b)] \cite{Atumi2013}. Although the theory also discussed how the N is located in the $\alpha$ layer [Fig. \ref{f1} (a)], the atomistic reason for the selective NV alignment in [111] \cite{Michl2014,Lesik2014,Fukui2014} is still unknown.

In this study we address this question by using first-principles energetics. We found that the N incorporated at the kink site of the [$\bar{1}$$\bar{1}$2] step edge is embedded in the subsurface in a sequence of kink-flows along this step and finally becomes the [111]-oriented NV center. Our theoretical scenario qualitatively explains the experimental results \cite{Michl2014,Lesik2014,Fukui2014}. 

We assumed that the N atoms participating in the formation of NV centers are supplied from the vapor phase during CVD. This assumption is justified by the experimental observation that the NV centers are aligned in the CVD growth directions for all (100) \cite{Pham2012}, (110) \cite{Edmonds2012} and (111) substrates \cite{Michl2014,Lesik2014,Fukui2014}. Hence, we studied the behavior of an external N atom attached to C(111) growing in the step-flow mode. The step-down direction was [$\bar{1}$$\bar{1}$2] in accordance with structure identification experiments for the steps on C(111) \cite{Tsuno1994,Tokuda2014}. A kink was  introduced as a terminator of the step edge because these experiments \cite{Tsuno1994,Tokuda2014} also show that the lateral step-flow growth of C(111) in CVD proceeds via kink flow. We did not consider the case of N incorporated at the $\beta$ carbon layers, because the displacement of N to the $\beta$ site, once incorporated in the $\alpha$ layer of the kink \cite{Fukui2014}, should be negligible before the subsequent carbon deposition.  

\begin{figure}[t]
\begin{center}
\includegraphics[width=7cm,bb=0 0 529 250]{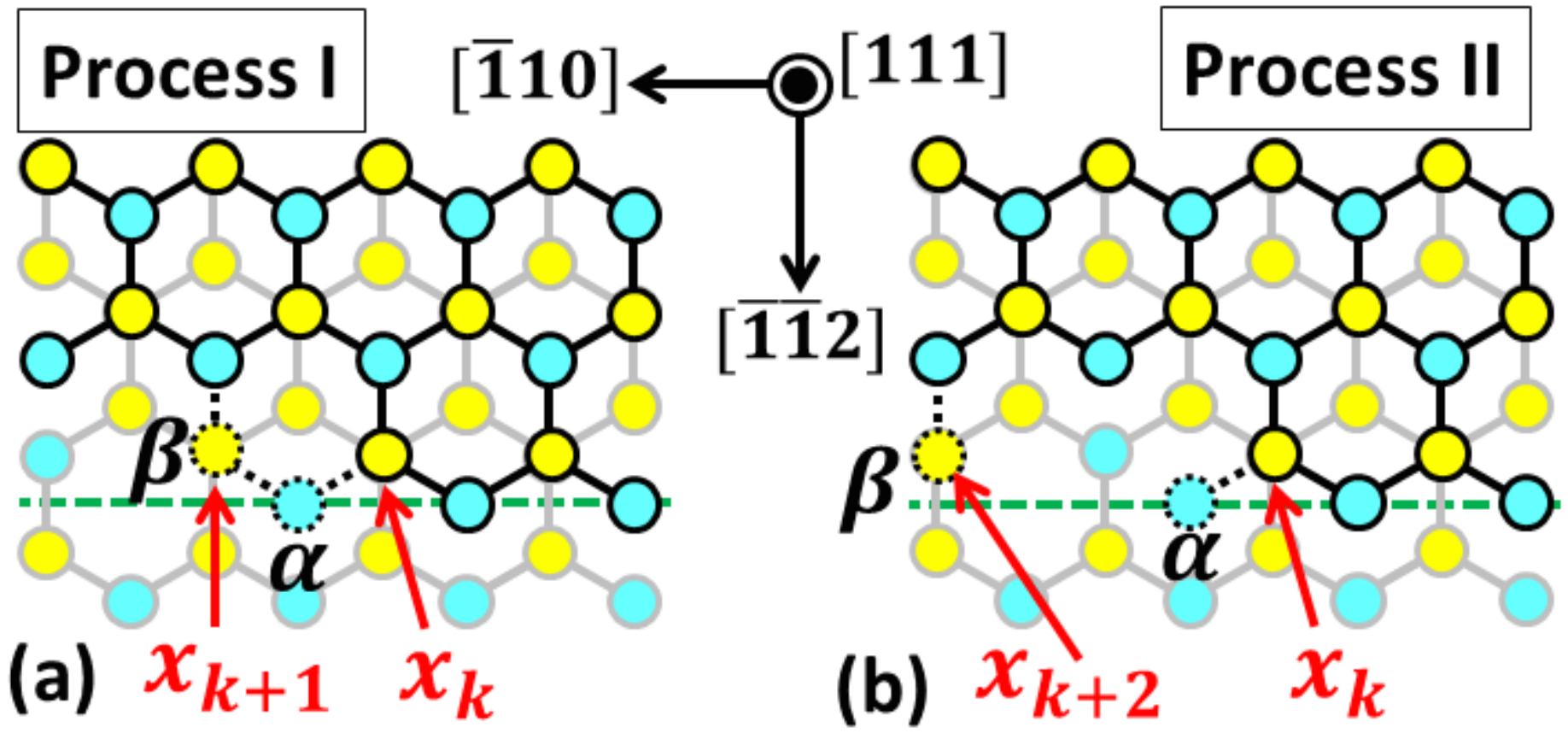}
\end{center}
\caption{Atomistic kink flow processes on C(111) considered in this study. Blue and yellow circles represent carbon atoms in the $\alpha$ and $\beta$ sublattice layers, respectively. Two carbon bilayers are shown from the topmost surface; the first and second bilayers are represented in dark and light colors, respectively. The H atoms are omitted for clarity. See note \cite{hydro} for how the surface is hydrogenated in the calculation. The dashed green line segments are visual guides to show the [$\bar{1}$10] direction of the step edge along which the kink flow occurs. The step-down direction is [$\bar{1}$$\bar{1}$2] for both panels. (a) Process I, in which the kink terminated at $x_{k}$ grows to $x_{k+1}$ by absorbing two C atoms. (b) Process II, in which a C atom occupies the $\beta$ site at $x_{k+2}$ and leaves a vacancy at $x_{k+1}$. }
\label{f2}
\end{figure}

An example of the kinked step edges considered is shown in Fig. \ref{f2}. The kink flow along the step edge arises from the movement of the kink site by the successive attachment of carbon atoms to the kink. The step-flow growth of C(111) is in units of bilayers \cite{Tsuno1994,Battaile1997,Battaile1998,Tokuda2014}, which are double layers composed of $\alpha$ and $\beta$ layers [see Fig. \ref{f1} (a)]. Thus we restricted ourselves to considering this kink-flow growth mode that proceeds in units of two carbon atoms, one in the $\alpha$ layer and the other in the $\beta$ layer. According to the energetics, the $\beta$ carbon atom should form a kink terminal, shown as $x_{k}$ in Fig. \ref{f2} (a). 

We considered two elementary processes of kink flow (Fig. \ref{f2}).
In process I [Fig. \ref{f2} (a)], two carbon atoms (shown as circles bounded by dotted curves) are attached to the kink at $x_{k}$. Then the $x_{k+1}$ site forms a new kink site in order to minimize the kink energy. In process II [Fig. \ref{f2} (b)], a carbon atom occupies the $\beta$ site at $x_{k+2}$ while the other C atom is the same as that in Fig. \ref{f2} (a). The latter process creates a vacancy at the $\beta$ site at $x_{k+1}$ in the surface bilayer, which is critical to the formation of the NV center aligned in [111]. 

The C(111) surface structure was modeled  using a four-bilayer thick slab in a repeated supercell with 5$\times$4 periodicity in the surface directions. A vacuum layer about 1.1 nm thick was inserted to decouple the spurious interactions between the slab and its self-images. All the C atoms on the terrace were hydrogenated. For the C atoms at the step edge, we investigated both the hydrogen-rich and hydrogen-poor cases. 
The hydrogen-rich step edge was constructed by attaching two H atoms to each C atom in the step edge. Lower-terrace C atoms nearest to the step-edge were terminated with one H atom each. In the hydrogen-poor step edge, only one H atom was attached to each C step-edge atom and the corresponding lower-terrace C atoms were not hydrogenated except the one closest to the kink position, to which one H atom was attached.

\begin{figure}[t]
\begin{center}
\includegraphics[width=8.5cm,bb=0 0 678 494]{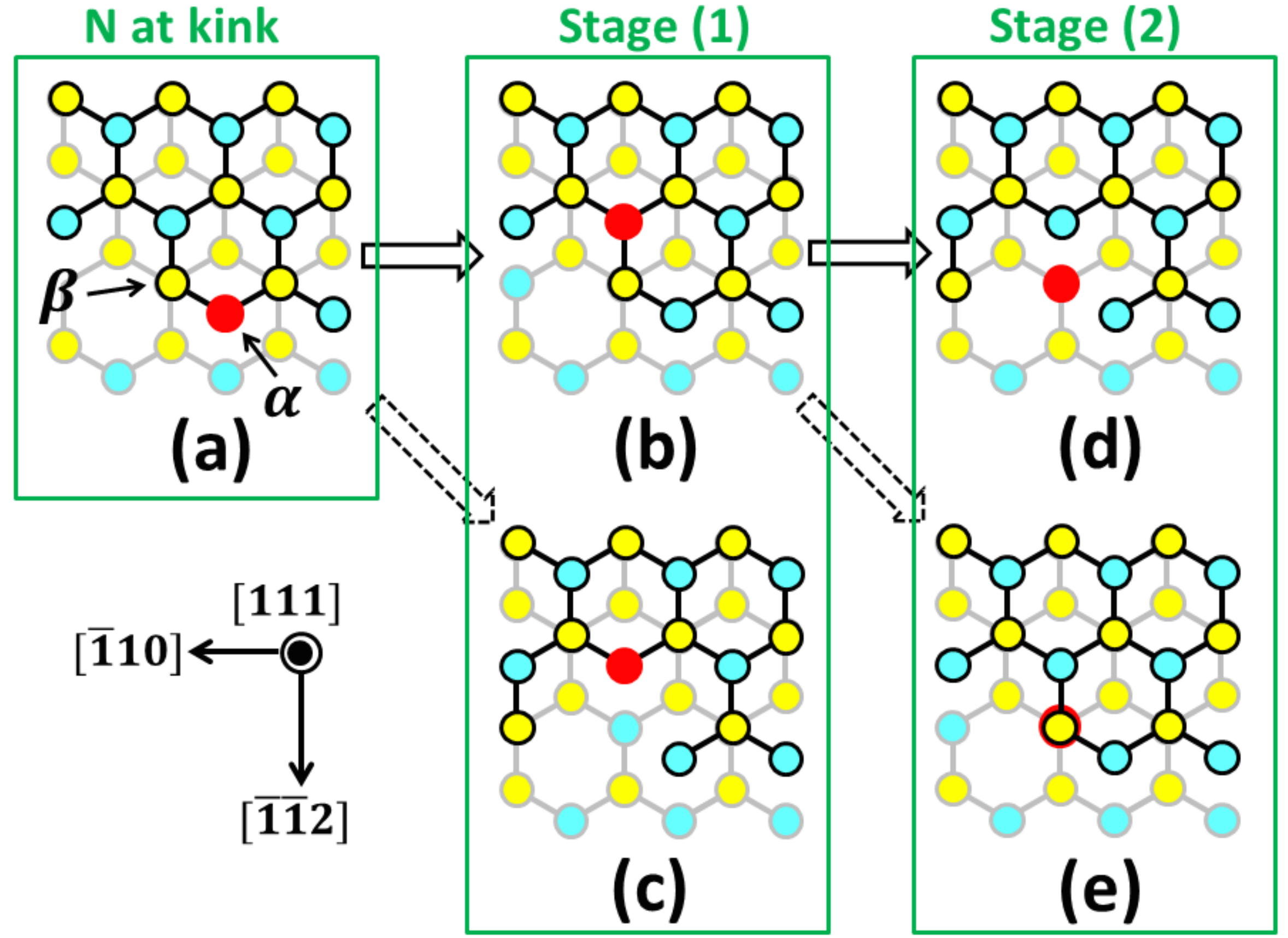}
\end{center}
\caption{Growth of [111]-aligned NV center in C(111). Red circles represent the N atoms. The other conventions for the atoms are the same as in Fig. \ref{f2}. (a) N at the $\alpha$ site of kink $\sim$0.5 eV lower in energy than the N at $\beta$, irrespective of the hydrogen content at the step edge \cite{Fukui2014}. Panels (b) and (c) are possible structures in stage (1). The N at the kink is about to be incorporated in the surface terrace via process I [(b)] and II [(c)]. Panels (d) and (e) are possible structures in stage (2). The N in the surface terrace is about to be overgrown via process II [(d)] and I [(e)]. 
}
\label{f3}
\end{figure}

To calculate the total energies of the N impurity in the surface, we performed first-principles electronic structure calculations \cite{QE} throughout this work based on density functional theory (DFT) \cite{HK,KS} within the generalized gradient approximation to the exchange-correlation energy \cite{PBE}. Further details of a calculation method are described in ref. \cite{method}. 
Because an NV center can exist in the negative (NV$^{-}$) and neutral (NV$^{0}$) charge states\cite{Doherty2013}, we optimized 
the structures of  
the NV centers in C(111) for both charge states.

The entry point for an externally supplied N atom into diamond is the $\alpha$ site at the kink, as shown in Fig. \ref{f3} (a). The preference for the $\alpha$ site over the $\beta$ site was theoretically identified in a previous study \cite{Fukui2014}. 

\begin{figure}[t]
\begin{center}
\includegraphics[width=8.5cm,bb=0 0 700 232]{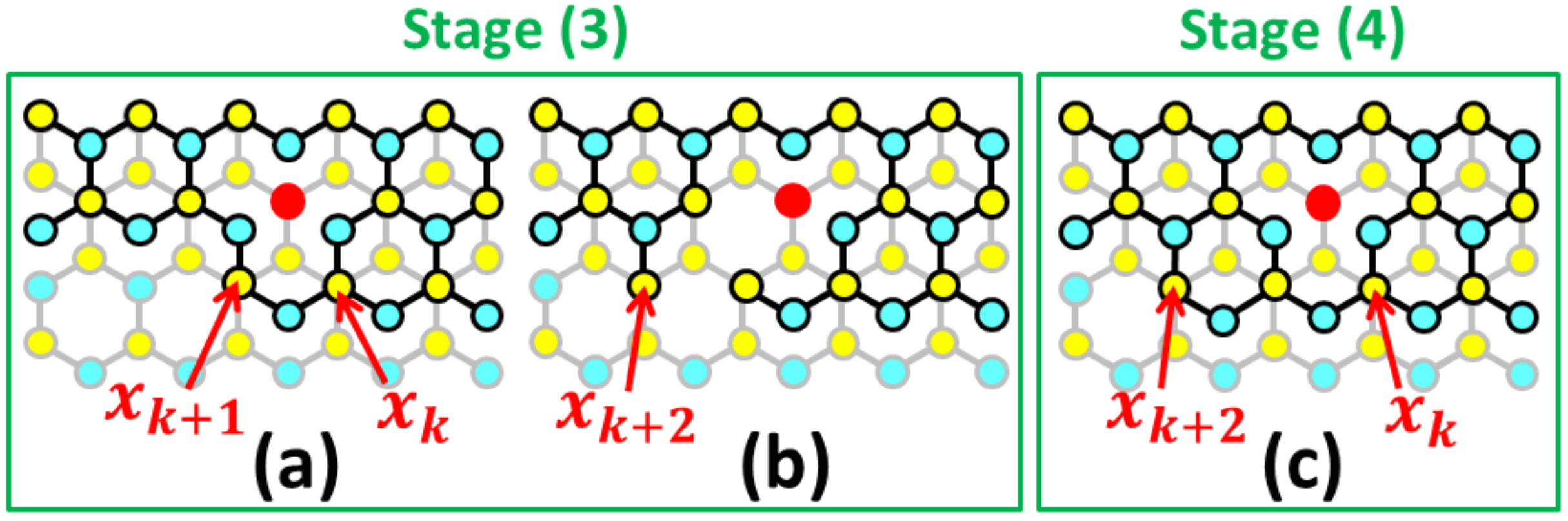}
\end{center}
\caption{(a), (b): A [111]-aligned NV center in stage (3) near completion. Red circles represent the N atoms. The other conventions for the atoms are the same in Fig. \ref{f2}. (a) The NV is surrounded by the carbon atoms in the surface bilayer through process I [Fig. \ref{f2} (a)]. (b) The structure that is degenerate with (a) in terms of energy when the step edge is hydrogen-poor. (c) The NV completely embedded in the C(111) surface, which is final stage (4). This configuration can be obtained from both (a) and (b) by adding two C atoms.}
\label{f4}
\end{figure}

The structures studied here are shown in Fig. \ref{f3} (b)--(e), Fig. \ref{f4} (a) and Fig. \ref{f4} (b). We classify these into three stages depending on the location of N with each stage having two branches. In stage (1), N is incorporated in the topmost surface terrace [Fig. \ref{f3} (b) and Fig. \ref{f3} (c)], and then in stage (2)  the C bilayer is overgrown [Fig. \ref{f3} (d) and Fig. \ref{f3} (e)]. In stage (3), the NV is embedded in the second bilayer [Fig. \ref{f4} (a) and Fig. \ref{f4} (b)]. In stage (4) [Fig. \ref{f4} (c)], the NV is separated from the step edge. 
It should be noted that we are considering the case where the position of N is not changed but the kink and step edge move toward [$\bar{1}$10] and [$\bar{1}$$\bar{1}$2], respectively, through the addition of the carbon atoms traveling from the vapor phase. 

Comparing the total energies of the two structures in each stage (Table \ref{t1}) indicates that the [111]-aligned NV center can grow via the structural evolution shown in Fig. \ref{f3} (a) $\rightarrow$ Fig. \ref{f3} (b) $\rightarrow$ Fig. \ref{f3} (d) $\rightarrow$ Fig. \ref{f4} (a). In stage (1), the N atoms [Fig. \ref{f3} (b)  and Fig. \ref{f3} (c)] are about to be embedded in the surface terrace as a result of adding two carbon atoms to the kink at $x_{k}$ via process I [Fig. \ref{f3} (b)] and II [Fig. \ref{f3} (c)]. The energy of the structure in Fig. \ref{f3} (b) is much lower than that of Fig. \ref{f3} (c), irrespective of the hydrogenation content at the step edge (Table \ref{t1}). Thus, the lateral step-flow growth arising from the kink flow can embed the N atom at the step edge into the surface just like the step flow of pure carbon bilayers.

\begin{table}[t]
\caption{The total energies of the structures in Fig. \ref{f3} (b), Fig. \ref{f3} (d) and \ref{f4} (a) relative to those in Fig. \ref{f3} (c), Fig. \ref{f3} (e) and \ref{f4} (b), respectively.}
\label{t1}
\begin{tabular}{ccccc}
\hline
   & \multicolumn{2}{c}{H-rich step} &  \multicolumn{2}{c}{H-poor step} \\
 \cline{2-5} 
   & NV$^{0}$  & NV$^{-}$ & NV$^{0}$  & NV$^{-}$ \\
 \cline{2-5}
  & \multicolumn{2}{c}{(eV)} & \multicolumn{2}{c}{(eV)} \\ 
 \hline
Stage (1) [Fig. \ref{f3}(b)] & $-$2.49 & $-$2.41 & $-$1.29 & $-$1.21 \\
Stage (2) [Fig. \ref{f3}(d)] & $-$1.62 & $-$1.63 & $-$1.42 & $-$1.42 \\
Stage (3) [Fig. \ref{f4}(a)] & $-$0.49 & $-$0.23 & $-$0.01 & $+$0.77\\
\hline
\end{tabular}
\end{table}

Stage (2) [Fig. \ref{f3} (d) and Fig. \ref{f3} (e)] explains how the vacancy is created on top of the N atom to produce the NV center aligned in the [111] direction. Here, process II occurs and contributes to the formation of the [111]-aligned NV center. The geometries in stage (2) illustrate the location in which the N atoms, now incorporated in the terrace, are about to be overgrown by additional carbon bilayers. The structure of Fig. \ref{f3} (d) corresponds to process II in the attachment of two carbon atoms to the kink at $x_{k}$. The total energy of this structure in both charge states is $\sim$1.5 eV lower on average than its counterpart in Fig. \ref{f3} (e), generated by process I, insensitive to the hydrogenation level of the step edge (Table \ref{t1}). The results for stage (2) arise from the stability of the electron lone pair of the N atom in the surface terrace that is threefold coordinated. 

The NV center is then grown in the C(111) surface in stage (3) [Fig. \ref{f4} (a) and Fig. \ref{f4} (b)]. The total energy difference between the structures in Fig. \ref{f4} (a) and Fig. \ref{f4} (b) shows that the completion of the NV center in  process I of the kink flow may be favored when the step edge is hydrogen-rich. For the hydrogen-poor step edge, the two structures are close in energy for NV$^{0}$ and the structure in Fig.\ref{f4} (b) is favored for NV$^{-}$ (Table \ref{t1}). Thus, a high hydrogen chemical potential is necessary to proceed to stage (4) [Fig. \ref{f4} (c)], in which the NV center is created.

The atomistic mechanism that we have described above delivers several messages. (i) V is  closer to the vacuum than N, as shown in Fig. \ref{f1} (a). The lone-pair orbital of an N atom placed in the $\alpha$ layers, on which a V is created, points into the vacuum. Experimental data show that a majority of the NV centers have this structure \cite{Michl2014}. Combining the C(111) bilayer structure and the energy gain arising from the  reduction of the surface dangling bonds, the most achievable stacking order of carbon layers from the surface to the bulk during CVD is $\alpha$$\beta$$\alpha$$\beta$$\cdots$, where the N atoms are located in only the $\alpha$ layers. This is in accordance with a suggestion by a preceding first-principles study \cite{Atumi2013}.   
(ii) Hydrogen-rich environments are required for preferential growth of the [111]-aligned NV centers. The experiments in which the nearly perfect alignment of the NV centers occurs in the samples grown with the 1.5\% CH$_{4}$/H$_{2}$ ratio or less \cite{Lesik2014,Michl2014,Fukui2014} imply that as many of the surface carbon atoms as possible should be terminated with H atoms. Our calculation results are consistent overall with this requirement. The exception is that, in stage (3), the NV$^{-}$ defect at the H-poor step edge may occur, in which the V is not fully isolated in the terrace [Fig. \ref{f4} (b)], depending on the chemical potential of H. The grand potential of the H-rich (H-poor) step edge is lower than that of the H-poor (H-rich) one, if we define the hydrogen chemical potential as the free energy of H atoms (H$_{2}$ molecules) at a temperature of $\sim$1000$^{\circ}$C and a pressure of $\sim$150 Torr, similar to the experimental conditions \cite{Lesik2014,Michl2014,Fukui2014}. To refine the modeling resolution of these competing energetics further, it may be necessary to take account of the vibronic entropies of the substrate at finite temperatures and/or the deviation of the H-atom content from equilibrium that may occur in reactors by the  excitation of high-density H plasmas. Although both these refinements are beyond the scope of the present study, they are interesting future areas of research.
(iii) Throughout this study, our results suggest that the [111]-aligned NV centers are grown in the crystal during CVD, which is consistent with the argument based on experimental results \cite{Edmonds2012}. The growth scenario is further supported by a recent advanced DFT study that shows that V diffusion may induce the formation of divacancies (V$_{2}$) instead of NVs, and also the immobilization of V because of the  production of the V$^{-}$ defects \cite{Deak2014}. On the other hand, another recent first-principles study \cite{Karin2014} proposes that nearly 90{\%} of NV defects can be aligned in the [111] direction if the sample is under a bi-axial, 2{\%} compressive strain normal to [111] and thermally annealed at 970$^{\circ}$C. This supports the post-growth process for the pre-existing NV defects. It should be noted that the question of which of the alignment mechanisms is more feasible is unresolved at present, which is another intriguing  issue.

In conclusion, we have obtained an atomistic mechanism for the production of NV centers selectively aligned in the [111] direction of the C(111) surface. Starting from the N-substitution of the $\alpha$ site of the kink, N is incorporated in the surface during the lateral bilayer growth, leaving a vacancy attached to top of the N site. Our model is consistent with current experimental data.


We used the VESTA ver.3.1.1 software \cite{Momma2011} to draw the atomic models in Fig. 1.


\begin{thebibliography}{9}
\bibitem{method}
The ultrasoft pseudopotentials \cite{VDB,LAASN} were adopted to represent the interactions between the valence electrons and ionic cores of all C, N, and H atoms. The wavefunctions were expanded in the plane-wave basis set up to the cut-off energy of 50 Ry. The augmentation charges of the ultrasoft pseudopotentials were also expanded in plane waves up to the cut-off energy of 320 Ry. The self-consistent-field (SCF) loops were iterated until the total energies of consecutive SCF steps was less than 1.5$\times$10$^{-6}$ eV/supercell, where the forces were calculated. The structure optimization was performed until the absolute values of all the components of all forces were smaller than 3$\times$10$^{-3}$ eV/{\AA}.
\bibitem{hydro}
We assumed that the upper and lower terrace C atoms in the $\alpha$ layers were mono-hydrogenated. In panel (a), one H atom was attached to the C atom at $x_{k+1}$. In panel (b), we assumed that two H atoms were attached to the C atom at $x_{k+2}$, whereas the one below the vacancy at $x_{k+1}$ had an unpaired dangling bond.
\end{thebibliography}
\end{document}